\documentstyle[amssymb,aps]{revtex}

\begin{document}
\title{High field transport in strained Si/GeSi double heterostructure: a
Fokker-Planck approach}
\author{F. Comas\thanks{%
Permanent Address: Depto. de F\'{\i}sica Te\'orica, Univ. de la Habana,
Vedado 10400, Havana, Cuba} and Nelson Studart}
\address{Departmento de F\'{i}sica, Universidade Federal de S\~{a}o Carlos,\\
13565-905, S\~{a}o Carlos, SP, Brazil}
\maketitle

\begin{abstract}
We report calculations of high electric field transport for the case of a
strained Si/GeSi double heterostructure (DHS) considering transport along
the Si channel and by applying the analytical Fokker-Planck approach (FPA),
where the process is modeled as drift-diffusion in energy space. We limit
ourselves to electronic transport in the conduction band of the strained Si,
where an energy shift between the otherwise degenerate six energy valleys
characterizes the band alingment in the DHS. Intervalley phonon scatterings
are considered while intravalley acoustic phonon scattering is ignored,
leading to results valid for high enough temperatures. Our results are
compared to previous theoretical works where Monte Carlo simulations were
applied. A reasonable agreement between the two approaches is obtained in
the high electric field regime.

PACS: 72.10.Di; 72.20.Ht; 73.20.Dx;
\end{abstract}

\newpage

\section{Introduction}

Important developments have been taking place in the growth of strained
heterostructures based on Si and Ge$_{x}$Si$_{1-x}$ layers \cite{b1,b2}. The
particular case of Si layers pseudomorphically grown on relaxed Ge$_{x}$Si$%
_{1-x}$ buffers is rather interesting because of the large mobilities
reported, exceeding the corresponding bulk $Si$ values \cite{b3,b4}. It has
been shown that in this type of double heterostructures (DHS) the otherwise
degenerate six $\Delta _{6}$ valleys of bulk Si are shifted in energy. If
the DHS is grown along a high symmetry direction (say, the $<001>$
direction), the two valleys along this direction are shifted downwards in
energy, while the other four valleys are shifted upwards. As a result, we
are led to two $\Delta _{2}$ valleys with a bottom energy below the
corresponding one for the bulk valleys and four $\Delta _{4}$ valleys with a
higher bottom energy. The energy shift between the $\Delta _{2}$ and $\Delta
_{4}$ valleys is empirically estimated as $\Delta E=0.6\,x$ eV. In this kind
of DHS, a quantum-well like band alignment in the Si channel was
unexpectedly found in 1985\cite{b1,b2}, which is due to tensile strains in
Si and ensures high-mobility $n$ channel for the doped structure. High
mobilities are explained as a consequence of the low effective mass of the
carriers in $\Delta _{2}$ valleys and also because the scattering efficiency
of the carriers (by phonons and impurities) is reduced. The above mentioned
facts are of great importance for device performance and this kind of
structures seems to have a prominent future for applications in micro and
optoelectronics \cite{b2}.

Monte Carlo calculations of high electric field mobilities and drift
velocities for electronic transport along the Si channel of Si/Ge$_{x}$Si$%
_{1-x}$ were reported by various authors \cite{b5,b6,b7,b8} and also applied
in the study of modulation-doped field-effect transistor (MODFET) structures 
\cite{b9}. More recently, hole transport parameters have been analyzed for
this type of systems \cite{b10}. While the electronic transport properties
of low field, low temperature Si/GeSi heterostructures are more or less well
understood \cite{b11}, the situation is not the same for high electric
fields. For high field transport, the effects of size quantization are
usually negligible and we can work within three dimensional models for the
valley structure. In Ref. \cite{b7}, size quantization was introduced in a
10 nm Si channel, but the authors had to include up to six subbands in the $%
\Delta _{2}$ valleys. Strain effects are introduced through the splitting
energy $\Delta E$ between the $\Delta _{2}$ and $\Delta _{4}$ valleys, which
ranges from $0.1\,$eV to $0.4\,$eV. Such relatively large values of the
energy shift effectively reduce the intervalley phonon scatterings between
the valleys (in comparison with the unstrained Si).

In the present paper we report results for high electric field electronic
transport along the Si channel of a Si/GeSi DHS. We model the system in
close analogy with the above mentioned works, taking into account the energy
shift between the $\Delta _{2}$ and $\Delta _{4}$ valleys and neglecting
size quantization. Rather than Monte Carlo simulations, we apply an
analytical Fokker-Planck approach (FPA), which treats transport as an
energy-diffusion process in energy space. This approach was proposed a long
time ago \cite{b12,b13} as an alternative for the consideration of
high-field transport in semiconductors. The theory bears a semiclassical
nature and has been recently revisited, the general formalism being
discussed with further details \cite{b14,b15,b16}. As a test of the FPA in
the case of a well known semiconductor, the present authors applied the
theory to bulk Si, where both experimental data and Monte Carlo simulations
for the high field drift velocity were successfully reproduced \cite{b17}.
The current calculations can also be considered as a way of testing the FPA
in a somewhat different system, which is of present day interest in high
technology applications.

The FPA is applicable when the energy exchanges between the carriers and the
surrounding medium (crystal lattice + external field) can be assumed
quasicontinuous. This latter condition apparently invalidates the method for
highly inelastic scattering mechanisms (as is the case of carrier scattering
by optic and intervalley phonons). However, if the carrier energy is large
enough, the exchanged energy becomes certainly low compared with the energy
of the carriers, and this is the case of high field transport. Hence, the
FPA is assumed to be valid if the condition $E_{av}>>\hbar \omega $ is
fulfilled, where $E_{av}$ is the average carrier energy and $\hbar \omega $
the exchanged energy (say, the phonon energy). The FPA has the advantage of
being analytical, and, whenever it can be applied, saves computational time
and allows a more transparent physical interpretation. Of course, it cannot
compete in accuracy with Monte Carlo simulations. However, we have found
that the FPA leads to comparatively good results even in cases where several
scattering mechanisms should be taken into account \cite{b17}. In contrast
with Monte Carlo simulations, the analytical approach involves less
realistic models for the analysis of a concrete semiconductor. More details
about the applied model and the fundamental theory are given in the next
sections.

The paper is organized as follows. In Sec. II we briefly summarize general
theoretical aspects of our work, in Sec. III details of the calculations for
the Si/GeSi DHS are presented, in Sec. IV the results of our calculations
are shown and comparison with previous works is made.

\section{General Theory}

The FPA considers transport in the spirit of a drift-diffusion process in
energy space. A certain distribution function (DF), $f(E,t)$, is defined
which depends on the carrier energy $E$ and the time $t$, such that $%
f(E,t)N(E)$ gives the number of carriers at time $t$ having their energies
in the interval $[E,E+dE]$, while the function $N(E)$ represents the density
of states (DOS). The DF obeys the equation \cite{b12,b13,b14,b15,b16} :

\begin{equation}
\frac{\partial}{\partial t}f(E,t)+\frac{1}{N(E)}\frac{\partial}{\partial E}%
J(E,t)=0,  \label{e1}
\end{equation}
where

\begin{equation}
J(E,t)=W(E)N(E)f(E,t)-\frac{\partial }{\partial E}\left[ D(E)N(E)f(E,t)%
\right] .  \label{e2}
\end{equation}
In Eq. (\ref{e2}) $W(E)$ represents a certain ``drift velocity'' of the
carriers in energy space and in fact gives the rate of energy exchange of
the carriers with the surrounding medium, while $D(E)$ is a kind of
diffusion coefficient. Equation (\ref{e1}) has the form of a continuity
equation for the carrier ``motion'' in energy space and leads to the
conservation of the carrier ``flux''. For a many valley system it should be
applied to each of the valleys separately and it is worth to note that Eq. (%
\ref{e1}) does not describe intervalley couplings. The quantity $J(E,t)$,
given by Eq. (\ref{e2}), is thus the carrier current density in energy
space. Under steady-state conditions the DF is time independent. Moreover,
we must have $J(E)=0$. Hence, we are led to

\begin{equation}
\frac{\partial }{\partial E}\left[ D(E)N(E)f(E)\right] =W(E)N(E)f(E).
\label{e3}
\end{equation}

We assume the existence of a phonon bath in equilibrium at the temperature $%
T $. The carriers interact with the phonons and the applied dc electric
field $\vec{F}$. We suppose, as an approximation, that the continuous
exchange of phonons between the carriers and the phonon bath does not affect
the thermal equilibrium of the latter. The coefficients $W(E)$ and $D(E)$
are split as follows

\begin{equation}
D(E)=D_{F}(E)+D_{ph}(E)\quad ,\quad W(E)=W_{F}(E)+W_{ph}(E).  \label{e4}
\end{equation}
The label ``$F$'' (``$ph$'') denotes the electric field (phonon)
contribution to these coefficients. We are obviously neglecting the
intracollisional field effect. The explicit form of these coefficients is
obtained below.

Equation (\ref{e3}) has the simple solution

\begin{equation}
f(E)=\exp \left\{ \int \left[ \frac{W_{ph}(E)}{D_{F}(E)}dE\right] \right\} ,
\label{e5}
\end{equation}
where $D_{ph}(E)$ was neglected. This approximation is very well fulfilled
in all the cases of interest for us \cite{b13,b14}. The practical usefulness
of the FPA lies in the possibility of the {\it analytical} performance of
the integral in Eq. (\ref{e5}).

Let us apply the above formalism to electrons in the $\Delta _{2}$ energy
valleys of strained Si of a Si/Ge$_{x}$Si$_{1-x}$ DHS. The $\Delta _{2}$
valleys are ellipsoids of revolution in $\vec{k}$-space with their
revolution axis along the $<001>$ crystallographic direction, coincident
with the growth direction of the DHS. To be specific, let us take $\vec{F}%
=(0,F,0)$, where the $x,y,z$ coordinates are taken along $<100>$, $<010>$
and $<001>$ respectively. The energy dispersion relation is given in the
form $\epsilon =\epsilon (\vec{p})$. In order to take into account
non-parabolicity of the band structure, we assume

\begin{equation}
\gamma (\epsilon )=\epsilon (1+\alpha \epsilon )=\frac{p_{t}^{2}}{2m_{t}}+%
\frac{p_{l}^{2}}{2m_{l}}=\frac{p^{\ast 2}}{2m_{0}}.  \label{e6}
\end{equation}
In Eq. (\ref{e6}) $m_{t}$ ($m_{l}$) is the transverse (longitudinal)
effective mass of the bulk Si conduction band electrons. We suppose the
masses are not changed by the strains in the Si layer: $m_{t}/m_{0}=0.19$
and $m_{l}/m_{0}=0.916$. Nonparabolicity is estimated also the same as in
bulk Si: $\alpha =0.5$ (eV)$^{-1}$. The Herring-Vogt transformation was
applied in writing the right-hand side (RHS) of Eq. (\ref{e6}), where $%
p_{t}=(m_{t}/m_{0})^{1/2}p_{t}^{\ast }$, $p_{l}=(m_{l}/m_{0})^{1/2}p_{l}^{%
\ast }$ and $p^{\ast 2}=p_{t}^{\ast 2}+p_{l}^{\ast 2}$.

The explicit expressions for $D_{F}(E)$ and $W_{ph}(E)$ can be directly
taken from Refs.\cite{b13,b14,b15}. Hence, for $D_{F}(E)$ we have

\begin{equation}
D_{F}(E)=\frac{2e^{2}F^{2}}{3m_{t}}\tau (\epsilon )\gamma (\epsilon
)/(\gamma ^{\prime }(\epsilon ))^{2},\quad  \label{e11}
\end{equation}
where $\gamma ^{\prime }(\epsilon )$ denotes the first derivative of the
function (defined in Eq. (\ref{e6})). For $\tau (\epsilon )$ we shall
consider one ``$g$'' intervalley phonon responsible for transitions between
the equivalent $\Delta _{2}$ valleys. Then the relaxation time reads as

\[
\frac{1}{\tau _{g}(\epsilon )}=C_{g}\left[ n_{g}(T)\sqrt{\gamma (\epsilon
+\hbar \omega _{g})}|1+2\alpha (\epsilon +\hbar \omega _{g})|\right. 
\]
\begin{equation}
\left. +(n_{g}(T)+1)\sqrt{\gamma (\epsilon -\hbar \omega _{g})}|1+2\alpha
(\epsilon -\hbar \omega _{g})|\theta (\epsilon -\hbar \omega _{g})\right] ,
\label{e12}
\end{equation}
where $\theta (\epsilon )$ is the step-function,

\begin{equation}
n_{g}(T)=\left[ \exp (\hbar \omega _{g}/k_{B}T)-1\right] ^{-1},  \label{e13}
\end{equation}
and

\begin{equation}
C_{g}=\frac{m_{d}^{3/2}D_{g}^{2}}{\sqrt{2}\pi \rho \hbar ^{3}\omega _{g}}.
\label{e14}
\end{equation}
In Eq. (\ref{e14}) $\rho $ is the semiconductor mass density, $\omega _{g}$
and $D_{g}$ are the phonon frequency and deformation-potential (DP) constant
respectively for intervalley phonons of type $``g"$. As a simplified model
for our calculations, we have considered just one $``g"$ phonon with an
energy $\hbar \omega _{g}=0.031\,$eV and DP coupling constant $%
D_{g}=11\,\times \,10^{8}$ eV/cm, which approximately corresponds to the
average of the three $``g"$ phonons of bulk Si reported in Table VI of Ref. 
\cite{b18}. Other bulk Si parameters are also taken from the same reference.

For ellipsoidal $\Delta _{2}$-valleys of the Si layer CB, the DOS is given by

\begin{equation}
N(E)=\frac{Vm_{d}^{3/2}}{\sqrt{2}\pi ^{2}\hbar ^{3}}\sqrt{E(1+\alpha E)}%
|1+2\alpha E|,  \label{e10}
\end{equation}
where nonparabolicity is taken into account and $m_{d}=\sqrt[3]{%
m_{t}^{2}m_{L}}$. In calculating $W_{ph}(E),$ we use the expression

\begin{equation}
W_{ph}(E)=\hbar \omega \left[ 1/\tau _{abs}(\vec{p})-1/\tau _{em}(\vec{p})%
\right] ,  \label{e9}
\end{equation}
where ``{\it abs}'' (``{\it em}'') denotes phonon absorption (emission) by
the electron. We must remember that

\begin{equation}
1/\tau (\epsilon )=1/\tau _{abs}(\epsilon )+1/\tau _{em}(\epsilon ),
\label{e15}
\end{equation}
the ``{\it abs}'' (``{\it em}'') term in Eq. (\ref{e15}) corresponds to the
first (second) term at the RHS of Eq. (\ref{e12}) for the case of
intervalley ``$g$'' phonons. Hence, using Eq. (\ref{e9}) and latter
expressions for $\tau _{abs}$ and $\tau _{em}$, we can find the explicit
form of $W_{ph}^{g}$.

Intravalley optical phonons do not contribute to transition rates because
the corresponding transitions are forbidden by the selection rules. The
contribution of intravalley acoustic phonons will be ignored in the present
work. For high temperatures and high carrier energies they should provide a
weak contribution to transport parameters and we shall limit ourselves to
this case.

As was remarked above, intervalley phonon scattering, strictly speaking, is
beyond the scope of the Fokker-Planck equation in its standard form [Eq. (%
\ref{e1})]. After the absorption or the emission of an intervalley phonon,
the electron is dropped away from one valley into another one, and such
abrupt transitions are not explicitly assumed in Eq. (\ref{e1}), where the
electron number is conserved within each valley. However, when we have
equivalent valleys (as is the case of $\Delta _{2}$ valleys), we can still
apply the Fokker-Planck equation in its standard form; electrons dropped
away from one valley fall into an equivalent one and vice versa, all happens
as the electrons always remained within the given valley. Of course, if the
valleys are not equivalent this argument no longer applies. This is the case
of electron transitions induced by the ``$f$'' intervalley phonons between
valleys $\Delta _{2}$ and $\Delta _{4}$, which are not equivalent in the
strained Si layer of the Si/GeSi DHS. For such intervalley electronic
transitions we must consider a different approach.

In order to estimate the effect of electron intervalley transitions due to
the interaction with ``$f$'' intervalley phonons we must realize that this
effect is actually weak and should introduce just small changes into our
final results. As for the case of ``$g$'' intervalley phonons, we assume
just one type of ``$f$'' phonons with energy $\hbar \omega _{f}=0.042\,$eV
and DP coupling constant $D_{f}=4\,\times \,10^{8}$ eV/cm (we again
estimated these parameters as averages from the three corresponding ``$f$''
phonons reported in Table VI of Ref. \cite{b18}). In all cases we considered
zero-order intervalley phonons in our model (a more realistic model must
consider first order intervalley phonons in the way discussed in Ref. \cite
{b19}). Electron transitions due to the interaction with ``$f$'' intervalley
phonons from $\Delta _{2}$ valleys up to $\Delta _{4}$ valleys is formally
switched on at $t=0$ and the DF in each of the $\Delta _{2}$ valleys evolves
in time by the approximate law $\frac{df}{dt}=-f/\tau _{f}$ , where $\tau
_{f}$ and the corresponding expressions for $n_{f}(T)$ and $C_{f}$ are
obtained from Eqs.(\ref{e12}), (\ref{e13}) and (\ref{e14}) by means of the
formal substitution: $g\rightarrow f$ and $\epsilon \rightarrow \epsilon
-\Delta E$.

After a time $\tau $ has elapsed, we obtain the result

\begin{equation}
f(E,\tau )=f(E)\exp \left[ -\tau /\tau _{f}(E)\right] ,  \label{e20}
\end{equation}
where $f(E)$ represents the DF without taking the $\Delta _{2}$-$\Delta _{4}$
intervalley scattering into account. Of course, this is just a rough
estimation of the effect and we are actually neglecting the inverse
intervalley process (from $\Delta _{4}$ to $\Delta _{2}$ valleys). Due to
the weakness of the effect, we shall assume that this procedure is
satisfactory. For the parameter $\tau $ in Eq. (\ref{e20}), we make the
reasonable estimate

\begin{equation}
\tau =\frac{1}{eF}\left[ \frac{m_{t}\hbar \omega _{f}}{2n_{f}(T)+1}\right]
^{1/2}.  \label{e21}
\end{equation}

\section{Calculation of the distribution function}

From Eq. (\ref{e5}), with the explicit application of Eqs. (\ref{e11}) and (%
\ref{e9}), we are led to the DF for $\Delta _{2}$ valleys without the
consideration of the $\Delta _{2}$-$\Delta _{4}$ intervalley scattering
process. The final result is

\begin{equation}
f_{j}(E)=E^{A_{j}}(1+\alpha E)^{B_{j}}\exp (\beta _{t}EP_{j}(E,T)),\quad
j=1,\,2,  \label{e22}
\end{equation}
where $A_{j}$ and $B_{j}$ are parameters (dependent on $T$ and $F$) and $%
P_{j}(E,T)$ is a polynomial in $E$. This structure is far from the
Maxwellian one. The DF describes a stationary non-equilibrium configuration
where an electron temperature $T_{e}$ cannot be defined. In Eq. (\ref{e22})
we shall measure all energies in units of $\hbar \omega _{g}$ and

\begin{equation}
\beta _{t}=\frac{3m_{t}m_{d}^{3}D_{g}^{4}}{4\pi ^{2}\rho ^{2}\hbar
^{4}e^{2}F^{2}}.  \label{e23}
\end{equation}
For $E<1$ we obtain

\begin{eqnarray}
A_{1} &=&\beta _{t}(n_{g}(T))^{2}(1+E_{0})(1+2E_{0})^{2}  \nonumber \\
B_{1} &=&\beta _{t}(n_{g}(T))^{2}(E_{0}-1)(E_{0}-1/2)^{2}  \nonumber \\
P_{1}(E,T) &=&(n_{g}(T))^{2}\sum_{i=0}^{4}a_{i1}E^{i},  \label{e24}
\end{eqnarray}
where the $a_{i1}$ are coefficients explicitly dependent on $E_{0}=\hbar
\omega _{g}\alpha $. For $E>1$ the following result is obtained

\begin{eqnarray}
A_{2} &=&\beta _{t}\left[
(n_{g}(T))^{2}2(1+8E_{0}^{2})-(2n_{g}(T)+1)(4E_{0}^{3}-8E_{0}^{2}+5E_{0}-1)%
\right]  \nonumber \\
B_{2} &=&\beta _{t}\left[
(n_{g}(T))^{2}2(8E_{0}^{2}+1)+(2ng(T)+1)(4E_{0}^{3}+8E_{0}^{2}+5E_{0}+1)%
\right]  \nonumber \\
P_{2}(E,T) &=&\sum_{i=0}^{4}a_{i2}E^{i},  \label{e25}
\end{eqnarray}
where the coefficients $a_{i2}$ are now dependent on $T$ through $n_{g}(T)$.
All the coefficients $a_{ij}$ are given in Table I.

In the interval $E>1$ we must distinguish three subintervals: $1<E<\Delta
E-E_{f}$, $\Delta E-E_{f}<E<\Delta E+E_{f}$ and $E>\Delta E+E_{f}$, where $%
E_{f}=\hbar \omega _{f}/\hbar \omega _{g}$. In the latter two subintervals,
the DF shall be given by

\begin{equation}
f_{j}(E,\tau )=f_{2}(E)\exp \left[ -\tau /\tau _{f}(E)\right] \quad ,\quad
j=3,4,  \label{e26}
\end{equation}
in correspondence with our approach to intervalley transitions between $%
\Delta _{2}$ and $\Delta _{4}$ valleys.

Once we have determined the DF, we can calculate the electron average energy
by the expression

\begin{equation}
E_{av}=\int \left\{ Ef(E)N(E)\right\} dE/\int \left\{ f(E)N(E)\right\} dE.
\label{e27}
\end{equation}
In Eq. (\ref{e27}) we just consider $\Delta _{2}$ valleys, i.e., we assume
the $\Delta _{4}$ valleys essentially deprived of carriers, an approximation
that seems acceptable if the applied electric field is not extremely large.
Moreover, in Eq. (\ref{e27}) we should be careful in partitioning the
integration into the four energy subintervals mentioned above.

Another important quantity is the drift velocity $v_{d}$ given by

\begin{equation}
v_{d}=\frac{2eF}{3m_{t}}\int \left\{ \frac{\gamma (E)\tau _{g}(E)}{(\gamma
^{\prime }(E))^{2}}\left[ -\frac{df}{dE}\right] N(E)\right\} dE/\int \left\{
f(E)N(E)\right\} dE.  \label{e28}
\end{equation}
In Eq. (\ref{e28}) we must partition the integration in the same way as in
Eq. (\ref{e27}). We just considered the drift velocity from electrons in the 
$\Delta _{2}$ valleys.

\section{Discussion of results}

By direct application of the theory developed in the foregoing sections, we
have made numerical calculations of the average electron energy $E_{av}$ and
the electron drift velocity $v_{d}$ as functions of the temperature $T$ and
dc electric field $F$. Our results are essentially valid for high
temperatures as far as intravalley acoustic phonon scattering was ignored.
As it was remarked in Sec. I, the FPA is applicable for high electric
fields, when the condition $E_{av}>>\hbar \omega $ fulfils \cite{b14,b15}.
In the examined Si/GeSi DHS, we take into account the relatively large
energy shift between valleys $\Delta _{2}$ and $\Delta _{4}$ and consider
that just the $\Delta _{2}$ valley is substantially populated by carriers.
Hence, intervalley transitions between valleys $\Delta _{2}$-$\Delta _{4}$
are assumed to be weak processes and treated within a relatively coarse
approximation. As it was discussed in the previous sections, such
transitions are actually beyond the scope of standard FPA. All numerical
parameters used in computations were taken from Ref.\cite{b18} and also
shown in Sec. III.

In Fig.~1 we show our results for the average electron energy $E_{av}$ as a
function of electric field $F$ for $T=300$ K. Two values of $\Delta E$ were
considered: $0.4$ and $0.1$ eV. As expected, the results show a rather weak
dependence on $\Delta E$. The involved carrier energies are large enough,
thus ensuring the applicability of the FPA. After comparison of Fig.~1 with
Fig.~1(a) of Ref. \cite{b5}, we can see that a reasonable agreement was
actually achieved. It is important to notice that just one fitting parameter
was applied in the $\beta _{t}$ of Eq. (\ref{e23}), which could be related
to the overlapping integral describing the electron-phonon scattering
probabilities. This overlapping integral was assumed unity in the general
formulas of Sec. II, but actually it differs from unity when intervalley
phonons are present. The linear behavior of $E_{av}$ for $F\lesssim 5$ kV/cm
should not be realistic because the FPA is not valid in the low-field regime
as discussed exhaustively in previous works. \cite{b14,b15,b16}

In Fig.~2 we present the drift velocity $v_{d}$ as a function of the
electric field for $T=300$ K. The continuous curve represents our
calculations, while the dots were taken from Fig.~2(a) of Ref. \cite{b15}.
We just considered the case $\Delta E=0.2$ eV. As a matter of fact, the
curves for different values of $\Delta E$ are almost coincident (in close
agreement with the results shown in Ref. \cite{b15}) as can be expected from
general physical grounds. For lower electric fields our results deviate from
those of Ref. \cite{b15}, a reasonable result taking into account that the
FPA is valid for high electric fields. However, for very high electric
fields ($F>80$ kV/cm) we again notice an increasing deviation from the Monte
Carlo results of Ref. \cite{b15}. The FPA is not able to describe the
saturation value of $v_{d}$ in the very high electric field region. On the
contrary, it is obtained a decreasing behavior of the drift velocity at a
rate that becomes much higher as higher is the field. In our present
treatment (as well as in that of Ref. \cite{b15}) negative differential
mobilities are out of question as far as the contributions of electrons from
the $\Delta _{4}$ valleys are not considered. As a final remark, we should
notice a very good agreement with the results of Ref. \cite{b15} for a wide
interval of electric fields.

In Fig.~3 we show the same plots as those of Fig.~2, but now for $T=77$ K.
From a simple glance at Fig.~3, it is obvious that for such a low
temperature the agreement between the FPA and the Monte Carlo results of
Ref. \cite{b15} (Fig.~2(b)) is much worse. As it was discussed before, we
have ignored intravalley acoustic phonons. Even though this is not a
limitation of the FPA itself, it is a difficult task to include them in the
calculations. Furthermore, at low temperatures, quantum effects should
become relevant and a quantum approach should be required. Hence, we stress
that the FPA\ results are reliable just for high temperatures.

In this work we have compared our results with those from Monte Carlo
simulations.\cite{b15} Other possible comparisons could be done. For
instance, we could compare with Fig.~3 of Ref.\cite{b8} or Ref.\cite{b6}.
But all these results are in accordance with themselves and nothing
essential would be added. Our results are also comparable to those of Ref. 
\cite{b7}, where size quantization was examined. As it was said before, size
quantization is of relevance when the carrier energies are low enough (this
is the case of low field transport as discussed in Ref. \cite{b11}), but for
the high carrier energies involved in high field transport, size
quantization becomes irrelevant. In the revised literature we have not found
available experimental data for this kind of system. As a general remark, we
conclude that the FPA leads to results in acceptable agreement with those of
Monte Carlo simulations in those intervals of temperature and electric
fields where it is supposed to be valid. This conclusion is true in spite of
the more or less coarse simplifications we have to face within the limits of
this method. However, if wider intervals of electric fields or more accurate
treatment is required, Monte Carlo simulations or some other equivalent
numerical procedure should be necessary.

\section{Acknowledgments}

We acknowledge financial support from Funda\c{c}\~{a}o de Amparo \`{a}
Pesquisa de S\~{a}o Paulo (FAPESP). F.C. is grateful to Departamento de
F\'{i}sica, Universidade Federal de S\~{a}o Carlos, for hospitality.

\bigskip

\newpage

\begin{center}
FIGURES
\end{center}

FIG. 1. Average electron energy ( in eV) as a function of the electric field
for $T=300$ K. Two values of $\Delta E$ are examined: $0.4$ and $0.1$ eV.

\medskip

FIG. 2. Drift velocity as a function of the electric field for $T=300$ K.
Our results are represented by the continuous while the dots were taken from
Monte Carlo simulations of Ref. \cite{b15}. We have set $\Delta E=0.2$ eV.

\medskip

FIG. 3. Same as in Fig. 2 for $T=77$ K. The dots are Monte Carlo results
from Ref. \cite{b15}. As one can see the agreement is now much worse.

\begin{table}[tbp]
\caption{Coefficients $a_{ij}$: $i=0, \cdots 4$ , $j=1,2$}
\begin{tabular}{|c|c|}
\hline
$4E_0(4E_0^3+8E_0^2+11E_0+3)+1$ & $%
8E_0(8E_0^2+3)+(2n_g+1)(-16E_0^4+32E_0^3-44E_0^2+12E_0-1)$ \\ \hline
$4E_0(8E_0^3+12E_0^2+7E_0+1)$ & $%
8E_0^2n_g^2(8E_0^2+7)+(2n_g+1)(8E_0^3-12E_0^2+7E_0-1)$ \\ \hline
$8E_0^2(2E_0+1)^2$ & $64E_0^3n_g^2-8E_0^2(2n_g+1)(4E_0^2-4E_0+1)$ \\ \hline
$8E_0^3(2E_0+1)$ & $32E_0^4n_g^2+8E_0^3(2n_g+1)(2E_0-1)$ \\ \hline
$16E_0^4/5$ & $-16E_0^4(2n_g+1)/5$ \\ \hline
\end{tabular}
\end{table}

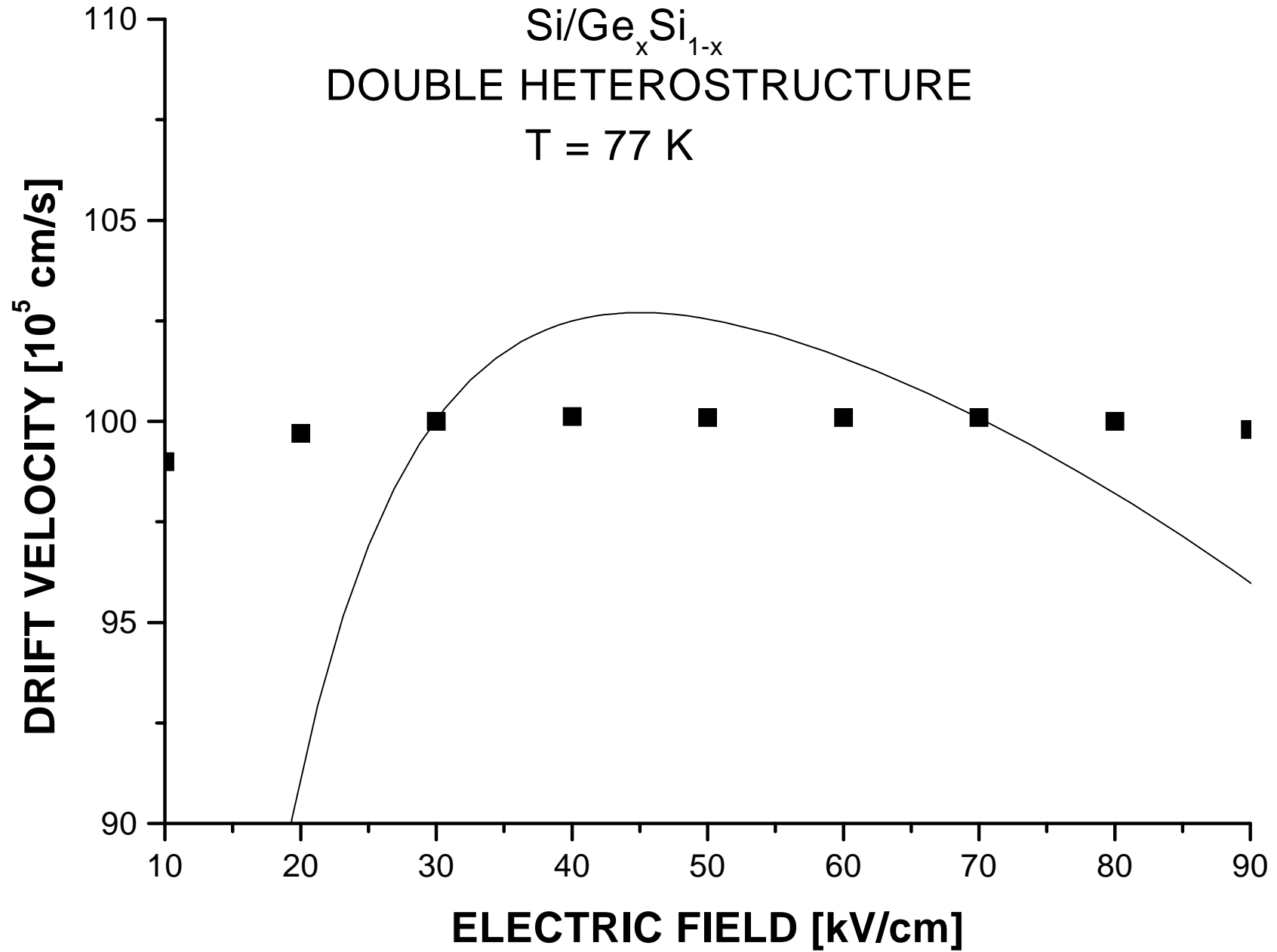
\end{document}